\documentclass[doublecol]{epl2} 
\usepackage{amssymb,amsmath}

\newcommand{\lgl}{\left\langle}
\newcommand{\rgl}{\right\rangle}

\newcommand{\bqu}{\begin{quote}}
\newcommand{\equ}{\end{quote}}

\newcommand{\eqa}{\begin{eqnarray}}
\newcommand{\eqe}{\end{eqnarray}}

\newcommand{\lk}{\left(}    \newcommand{\rk}{\right)}

\newcommand{\Lk}{\left\{}  \newcommand{\Rk}{\right\}}

\newcommand{\lK}{\left[} \newcommand{\rK}{\right]}

\newcommand{\br}{\mathbf r}%

\newcommand{\phibar}{\phi\kern-1ex\rule[1.25ex]{1ex}{.1ex}}

\newcounter{aufgnr}
\newcounter{aufgteil}

\title{Domain Walls in Helical Magnets: Elasticity and Pinning}
\shorttitle{Domain Walls in Helical Magnets: Elasticity and Pinning}
\author{T. Nattermann}
\shortauthor{T. Nattermann }

\institute{Institut f\"ur Theoretische Physik, Universit\"at zu K\"oln, Z\"ulpicher
Str. 77, D-50937 K\"oln, Germany\\
}
\pacs{75.10.-b}{General theory and models of magnetic ordering}
\pacs{75.60.-d}{Domain effects, magnetization curves, and hysteresis }
\pacs{75.70.Kw}{Domain structure (including magnetic bubbles and vortices)}

\abstract{
 Recently completely new types of domain walls (DWs) have been discovered in helical magnets,  consisting generically  of  a regular array of {\it pairs} of magnetic vortex lines \cite{Li+12}. 
 Only for special orientations DWs are free of vortices. 
 In this article we calculate their  elastic and pinning properties, using the pitch angle $\theta$ as a small parameter. 
In particular we show that vortex free DWs exhibit  long range elasticity which makes them very stiff and suppresses their pinning by impurities. 
 Their roughening transition temperature is of the order of the N\'eel temperature. 
 DWs including vortices (either by orientation or due to step formation above their roughening transition) show short range elasticity and strong pinning by impurities. 
 These results  apply both to centro-symmetric as well as to non-centrosymmetric systems. 
   The application to chiral liquid crystals 
      is briefly discussed. }

\begin{document}

\maketitle

\section{Introduction}
Pinning  plays a key role in condensed matter systems: 
it restores the state of zero resistance  in  type-II superconductors  by anchoring flux lines, it hardens  
steel by blocking the motion of dislocations \cite{Haasen77},   
but, in contrast,   prevents charge density waves to become  ideal conductors \cite{Frohlich54}.  
In ferroelectrics and ferromagnets pinning of domain walls influences  their coercivity  and  switching behavior  \cite{Hubert74,Kleemann07}, strongly relevant for potential applications as storage media \cite{Parkin+08}.
In all cases pinning and hysteresis
  result from the competition  of the impurity   potential, which favors  deformations of the condensed structure,  and   the   rigidity of the latter,  which penalizes them.  
  The appearance of a non-zero coercive force requires the emergence of { multistability} of the resulting effective potential landscape \cite{Brazovskii+04}. 
  
Recently a new type of magnetic DWs - different from Bloch or N\'eel walls - has been predicted  for helical magnets \cite{Li+12}.
Helical magnets are   abundant, occurring    as { metals} 
  and   alloys  \cite{Jensen+91,Lang+04,Pfleiderer+04,Uchida+06,Uchida+08}, { semiconductors} \cite{Kim+Seik+08}
  and  {multiferroics} \cite{Mostovoy06,Cheong+07,Chapon+04,Kimura+08,Tal+08,Yam+06,Chap+11,Arima11}.  
The latter group is most interesting for applications \cite{Cheong+07}.
It was shown in \cite{Li+12} that for almost all orientations  DWs   in these systems consist of a regular array of magnetic vortex lines. These walls can be driven by currents and,  in multiferroics, by electric fields
\cite{Li+12}.
Vortex walls   have indeed be seen in  circularly polarized X-rays in Ho \cite{Lang+04}, and by Lorentz-TEM in FeGe \cite{Uchida+08}. 
Only for special orientations DWs are vortex free.  Special cases of the latter have been studied  previously  by Hubert \cite{Hubert74} and will called  in thus article {\it Hubert walls} (see Fig.1). 
\begin{figure}[h]
\begin{center}
\includegraphics[height=4cm]{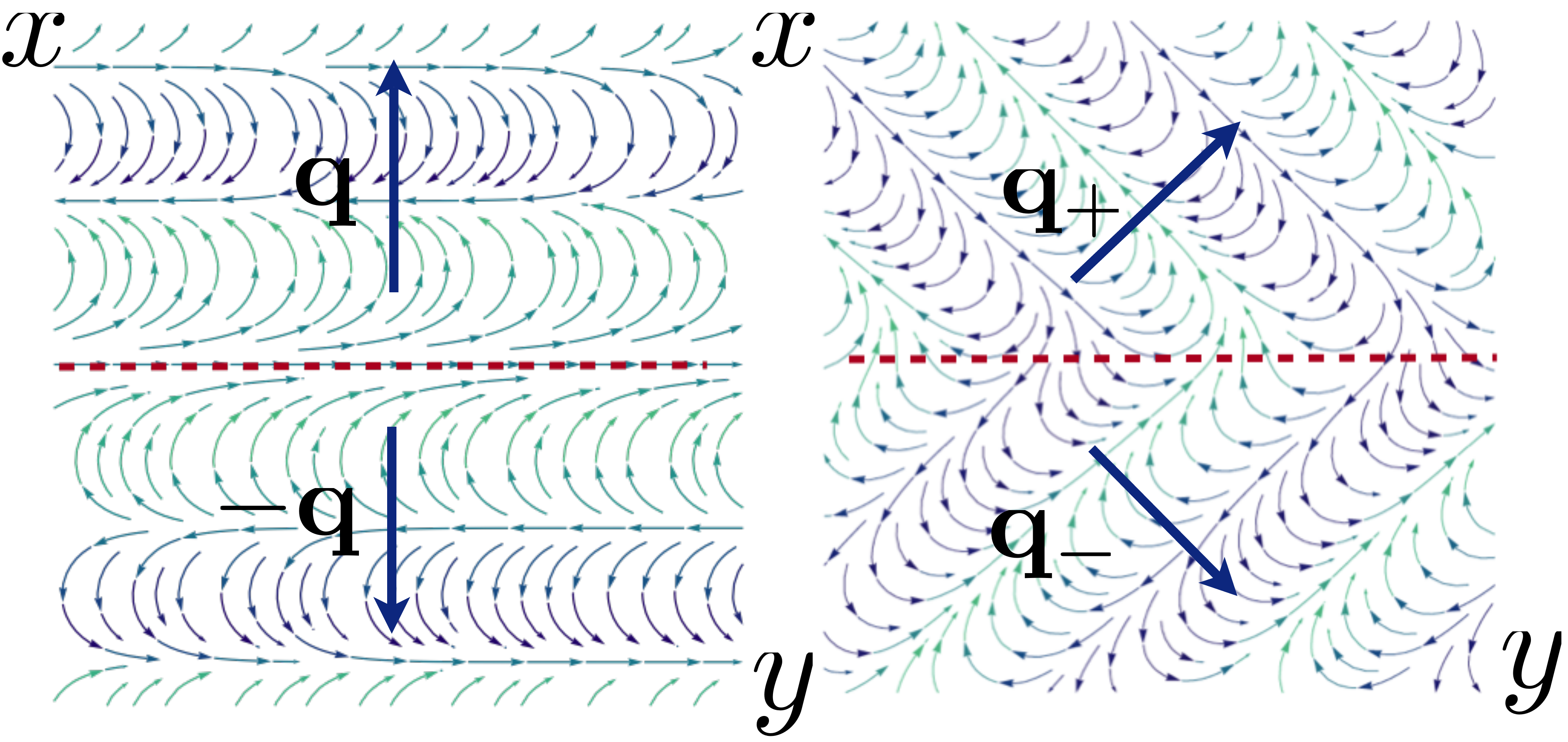}
\caption{Cross section of Hubert walls  in helical magnets. 
Left panel:  centrosymmetric system, right panel:  non-centrosymmetric system. 
The small arrows denote the orientation of ${\mathbf m}$, the large arrows the wave vectors $\bf q$. 
For systems where ${\mathbf m}\cdot\hat{x}=0$,  ${\mathbf m}$ has
been rotated by $\pi/2$ around $y$ for better visibility. 
}\label{figure1}
\end{center}
\end{figure} 

\section{Outline and results} 
In the present letter I investigate the elasticity and impurity pinning of DWs in helical magnets. 
I show that their micro-magnetic model  can be considerably simplified by writing it in a form similar to  the London theory of superconductors. 
In this description magnetic vortices result  from a fictitious magnetic field acting only inside the DW.   
Vortex free Hubert walls are found to  exhibit {\it non-local} elasticity which makes them too stiff to be pinned by impurities.
Hence they can easily disappear  from the sample, which 
  may explain why Hubert walls were not observed so far in experiment \cite{Weschke+04}.  
  Roughening of Hubert walls occurs by entropy or disorder driven  proliferation of steps. Steps  consist of { pairs} of vortex lines of the same vorticity which {\it attract} each other. 
The roughening transition temperature $T_R$ is calculated and found to be of the order of the N\'eel temperature.  Additional  disorder driven roughening of Hubert walls is shown to have a weak effect only.

  On the contrary, DWs including vortices exhibit local elasticity and are strongly pinned by   disorder. 
  I also discuss the effect of weak anisotropy on Hubert walls. Although most of the derivations are presented  for centrosymmetric systems,  it is shown that the results transfer to the non-centrosymmetric case as well. 
Finally I discuss some conclusions for DWs in chiral liquid crystals.
%
%
%
%

%
%
%
%
\section{Hamiltonian} 
To describe   helical magnets  I  use the appropriate micro-magnetic Hamiltonian  ${{\cal H}}[{\mathbf m}({\mathbf r})]$. 
${\mathbf m}=(m_x,m_y,0)$ denotes the magnetization, assuming $m_z=0$. 
Since in helical magnets both time and space inversion symmetry are broken, their paraphase  can be  either {\it centrosymmetric} or {\it non-centrosymmetric}. 

Centrosymmetry requires  invariance with respect to space and time inversion, i.e.  ${\mathbf r}\to-{\mathbf r}$ and ${\mathbf m}\to-{\mathbf m}$.
If there are only two modulation vectors, ${\mathbf q}=\pm(\theta/a)\hat x$, as in  most centrosymmetric   systems, one  finds  up  to quadratic terms in  ${\mathbf m}$ \cite{Hubert74}  
  \eqa\label{eq:Hamiltonian2}
    {\cal H}=\frac{J}{2}\int_{{\mathbf r}}\left[-\frac{\theta^2}{2a} (\partial_x {\mathbf m} )^2+ \frac{a}{4} (\partial_x^2{\mathbf m} )^2+ \frac{1}{a}({\mathbf \nabla}_{\perp}{\mathbf m})^2\right]. 
 \eqe
Here $\int_{{\mathbf r}}=\int d^3r$ and  ${\mathbf \nabla}_{\perp}={{\hat y}}\partial_y+{\hat z}\partial_z$. $\theta$ and $a$ denote the angle between adjacent spins along the x-direction and  the lattice constant, respectively.   
 The continuum approach is valid provided $\theta\ll\pi$. 
 Experimentally one finds $\theta\approx{0.27-0.73}$ under ambient conditions \cite{Jensen+91} and $\theta\to{0}$ under uniaxial pressure \cite{Andrianov+00}. 
In systems  where indirect RKKY exchange between 4f electrons \cite{deGennes62} results in  nearest neighbor ferromagnetic ($J>0$) and next nearest neighbor anti-ferromagnetic ($J'<0$)  interaction,  $\theta=\arccos(J/4|J'|)$ in the ground state \cite{Hubert74}.  

\smallskip 
 
 With the replacement $m_x+im_y=e^{i\phi}$ 
Hamiltonian (\ref{eq:Hamiltonian2}) can be written as 
  \eqa\label{mmHamiltonian-phi1}
{\cal H}=\frac{Ja}{8}\int_{\mathbf r}
\Lk\lK(\partial_x\phi)^2-q^2\rK^2+\frac{4}{a^2}(\partial_\perp\phi)^2+\lk\partial_x^2\phi\rk^2\Rk.
\eqe 
(\ref{mmHamiltonian-phi1}) includes a  quartic term $\lK(\partial_x\phi)^2-q^2\rK^2$, making the calculation of the elasticity of Hubert walls further below cumbersome.  
 I will  then resort  to an approximation and replace this expression   by $4q^2\lK\partial_x\phi-A({\mathbf r})\rK^2$. 
${A}({\mathbf r})$ is assumed to be constant inside a domain ($A=\pm{q}$) and to change smoothly from $- q$ to $q$  on a scale $q^{-1}$  
when crossing  the DW.   
The   resulting Hamiltonian  
\begin{eqnarray}\label{eq:Hamiltonian3}
{\cal H}_0=\frac{Ja}{2}\int_{{\mathbf r}}\left[q^2\left(\partial_x\phi-A\right)^2+\frac{1}{a^2}(\partial_{\perp} \phi)^2+\tfrac{1}{4}\lk\partial_x^2\phi\rk^2\right]
\end{eqnarray}
resembles the London theory of type-II superconductors. 
${{\mathbf  A}}=(A({\mathbf r}),0,0)$ plays the role of a vector potential which generates a fictive magnetic field ${\mathbf{B}}={\mathbf \nabla}\times{{\mathbf  A}}$  acting only inside the DW where it  creates vortices. $q^{-1}$ corresponds to the London penetration lengths. 

\section{Non-local elasticity of Hubert walls}
 Calculation of pinning forces requires the knowledge of the elasticity of  DWs \cite{Brazovskii+04}. 
 It is convenient to introduce a rotated coordinate system with one axis, say $\xi$, parallel to the (average) normal $\hat n$ of the DW and the two other axes, $(\eta, z)\equiv\bm\eta$, perpendicular to $\hat n$, i.e. parallel to the DW plane.  
Since the system is isotropic in the $yz$-plane, it is indeed sufficient to restrict  the normal $\hat n$ to  the $xy$-plane.  
 Rotation of $\hat n$ around the $x$ axis does not change the results. 
Thus  we define
\eqa
\xi=x\cos\alpha+y\sin\alpha, \quad\eta=-x\sin\alpha+y\cos\alpha\eqe with $\alpha$ the angle between the $x$ axis and  $\hat n$.  

\smallskip

The  energy of long wave length elastic DW distortions  $u(\bm\eta)$ from a planar reference configuration  then reads \begin{equation}\label{elastic1}
{\cal H}_{el}=\frac{1}{2}\int d{\bm\eta} d{\bm\eta}'{ \cal G}^{-1}({{\bm\eta}}-{{\bm \eta}}')u({{\bm \eta}})u({{\bm \eta}'}).
\end{equation}  
We begin with the Hubert wall where $\alpha=0$, $\xi=x,\eta=y$ and  hence ${{\bm \eta}}=(y,z)$ . 
To determine ${\cal G}({\bm\eta})$ I assume  
 that the DW distortions are of buckling type, i.e. free of vortices (but see below). 
The saddle point equation for the undistorted Hubert wall  
 can be solved { exactly} by writing  (\ref{mmHamiltonian-phi1}) as a $\psi^4$ theory, where $\psi=\partial_x\phi$ and $\psi(x\to\pm\infty)=\pm q$.   
 For an isolated planar wall the exact solution is    \cite{Hubert74}
 \eqa\label{phi-0}
 \phi_0(x)=\ln\cosh[q(x-x_0)].
 \eqe 
The choice $A({\mathbf r})=q\tanh (q{x})$ in (\ref{eq:Hamiltonian3}) reproduces (\ref{phi-0}). 
The corresponding solution 
 for a {\it distorted} Hubert wall is conveniently calculated from (\ref{eq:Hamiltonian3}) with the Ansatz   
\begin{equation}\label{deGennes}
\phi({\mathbf r})=\int_0^xdx'A(x')+\int_{\mathbf k} e^{i(k_yy+k_zz)-|{\mathbf k}||x|/\theta}\alpha_{{\mathbf k}},
\end{equation}
used before by Joanny and de Gennes in the context of contact lines \cite{Joanny+84}. Here $\int_{{\mathbf k}}=\int{dk_ydk_z}/(2\pi)^2$. The first term describes the unperturbed phase field, the second term its corrections up to a distance $|{\mathbf k}|^{-1}$ from the average wall position. 
 $\alpha_{{\mathbf k}}$  is to be determined from the condition  
 \eqa
 \phi\lK u(\bm\eta),\bm\eta\rK=0. 
 \eqe 
 To lowest order in $u$, one finds $\alpha_{{\mathbf k}}=-u_{{\mathbf k}}q$, provided $|{\mathbf k}|<q$.
Plugging (\ref{deGennes}) back into  (\ref{eq:Hamiltonian3})  
one obtains for the Fourier transform of ${\cal G}({\bm\eta})$ 
\begin{equation}\label{elasticenergy}
{\hat{\cal G}}_H^{-1}({\mathbf k})\approx J|q|^3\left( 4|{\mathbf k}|+a{\mathbf k}^2/3\right). 
\end{equation} 
(\ref{elasticenergy}) is valid for distortions $|qu|>1$. The dominant term $\sim|{\mathbf k}|$ results from the  long range self-interaction of the DW which a makes it very stiff. 
 The second term in (\ref{elasticenergy}) is the contribution from the increased  surface area due to the wall distortion and is relevant only for large ${\mathbf k}$. 
 In real space the energy expression (\ref{elasticenergy}) for the Hubert wall is {\it non-local}, 
$
{{\cal G}_H^{-1}({\mathbf r})\sim{J}(|q|/|{\mathbf r}|)^{3}}
$, 
 in contrast to Bloch and N\'eel walls, whose elasticity is  strictly local. As we will see this has dramatic consequences for the pinning of Hubert walls since a strictly planar wall is not pinned.

\section{Roughening transition of Hubert walls}
Next I consider the possibility of a roughening transition  \cite{Chaikin+95} of the Hubert wall which would render elasticity short range \cite{Noziere01}.  
A roughening transition occurs due to the formation and proliferation   of terraces, separated by steps,  on the   Hubert wall (see Fig.2).   
Steps can occur due to thermal fluctuations or disorder. 
To study step formation more in detail
 I consider a Hubert wall with a step at $y=0$, parallel to the $z$-axis.  
 The cross section of a step is shown in the left panel of Fig. 2. 
 The phase $\phi$ across the step can be   described by the function  
\begin{equation}
{\phi_s(x,y)\equiv\phi_0\left(x-{\tau\pi}/{q}\right)+\tau\pi \textrm{sign}\left( x-{\tau\pi}/{q}\right)}
\end{equation}
where ${\tau=\textrm{sign}y}$ and $\phi_0(x)$ is given in (\ref{phi-0}). 
${\phi_s(x,y)}$   is smooth across  $y=0$ for  $|x|\gg\pi/q$. 
On the contrary, in the region $|x|<\pi/q$  the \textrm{sign} of $\partial_x\phi_s$ is opposite for $y \lessgtr{0}$. 
Hence the integral  along a contour ${\cal C}$ inclosing the step gives $\oint_{\cal C}\phi=4\pi$, i.e. the step consists of {\it two} vortices. 
\begin{figure}[h]
\includegraphics[height=3.7cm]{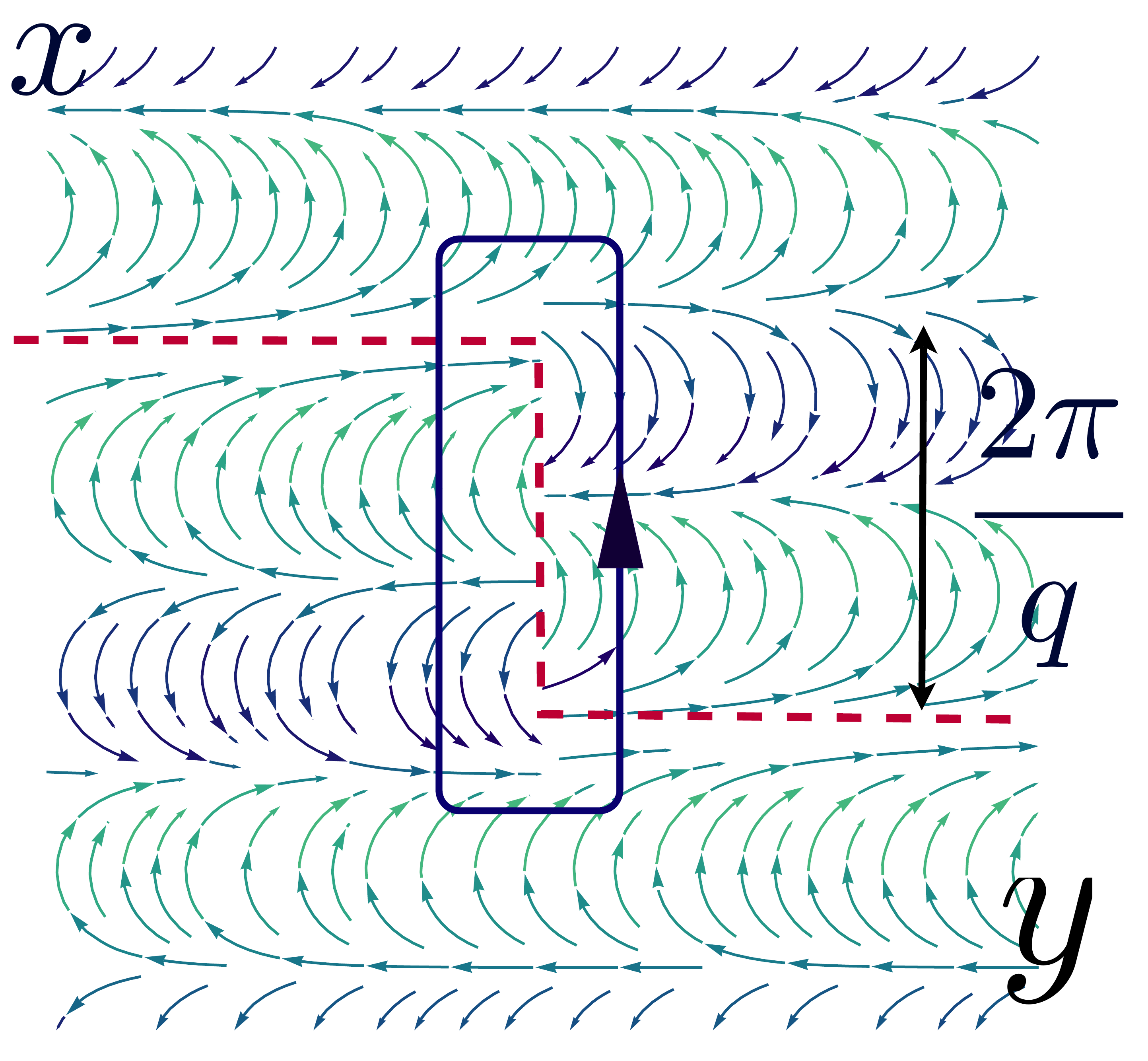}\includegraphics[height=3.4cm]{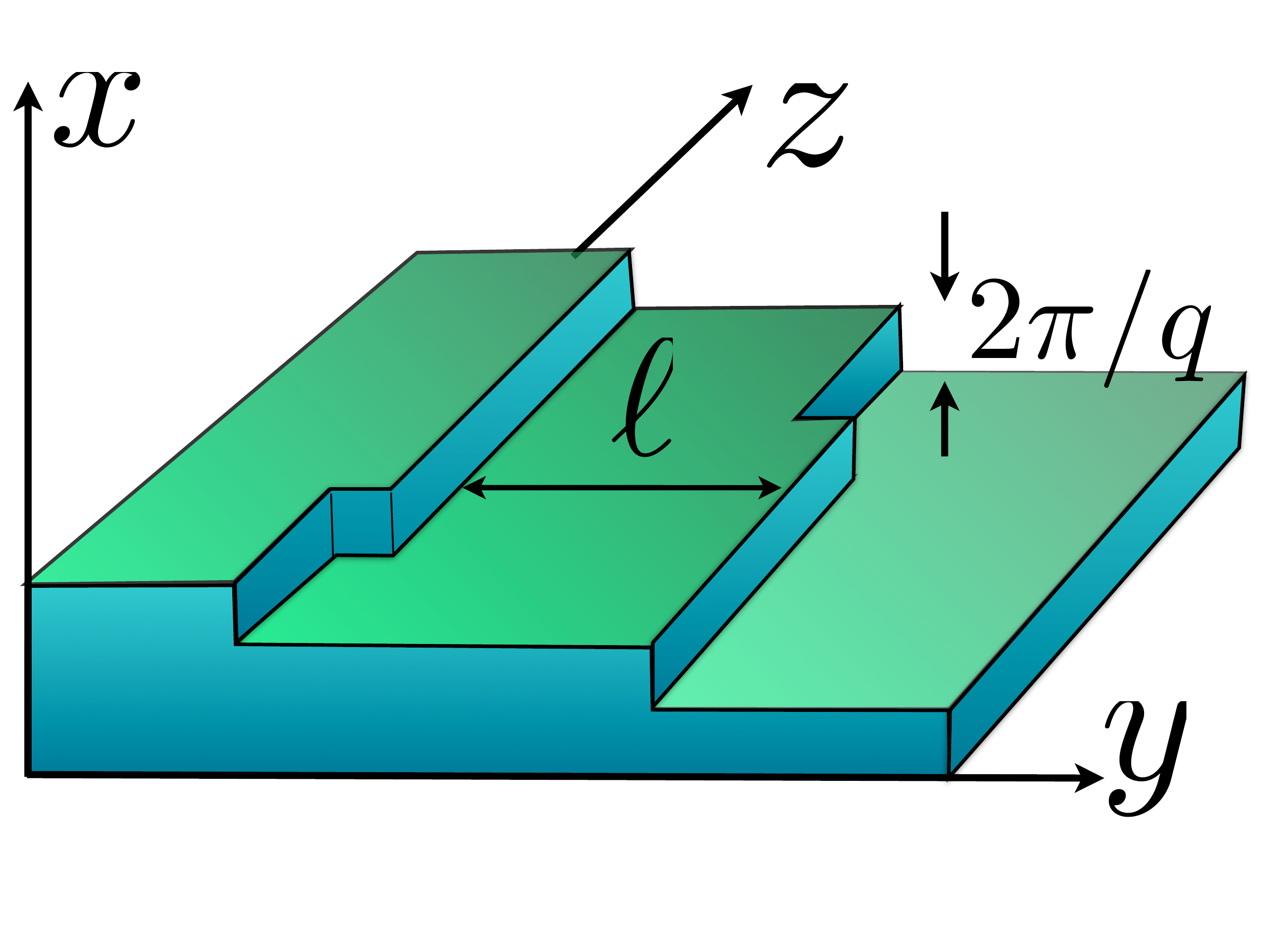}
 \caption{Left panel: Cross section of a  Hubert walls  with step before relaxation. 
The arrows denote again the orientation of ${\mathbf m}$. The step extends perpendicular to the picture plane. The contour ${\cal C}$ encloses two vortices. Right panel: Vortex wall consisting of an array steps  separated by terraces of Hubert walls.}\label{figure1}
\end{figure} 
In this construction the vortex configuration was restricted to a narrow slice of width $a$. The step energy per unit length, $\varepsilon_{s}$, is  of the order $J/(a\theta)$. 
Further relaxation of the configuration $\phi_s$ by allowing the vortex to extend over a larger region can only {\it decrease} $\varepsilon_{s}$. 
The step consists of two nearby vortices of the same vorticity. Each of them  fulfils  the  approximate saddle point equation 
\eqa\label{saddlepoint}
-\partial_x^4\phi_s+4\partial_\perp^2\phi_s=0,
\eqe
following from (\ref{mmHamiltonian-phi1})
provided the distance from the vortex centre is less than $2\pi/q$.  An approximate solution of (\ref{saddlepoint}) is given by the Ansatz 
\eqa\label{Ansatz}
\phi(x>0,y)=\arcsin\lK y/(\kappa^2x^4+y^2)^{1/2}\rK.
\eqe
 $\kappa$ is a variational parameter. With $x^2=ra\cos\varphi/\kappa,\,y=r\sin\varphi\,$,    (\ref{Ansatz}) gives for the vortex energy density $J{\cal E}_v(x,y)$
\begin{equation}\label{energydensity2}
 {\cal E}_v(x,y)
 \approx  \frac{1}{2ar^2} \lK \cos^2\varphi+\kappa^2\sin^2\varphi(1-4\cos^2\varphi)^2\rK. 
  \end{equation}
Minimization gives for  the vortex core energy per unit length  $E_v\approx0.6J/a$,   $\kappa=0.42$ \cite{Schenck+13} and hence
$ 
\varepsilon_s\approx 1.2 J/a. $ 
 The  two vortices attract each other  with a force  proportional to the length of the vortex line. 
 Using   the step energy $\varepsilon_{s}$ in the results for the roughening transition in the ASOS model \cite{Hasenbusch+97} one obtains for the roughening transition temperature of the Hubert wall
$ 
T_{R}^{(H)}\approx J/k_B. 
$ 
Adopting for the N\'eel temperature $T_N$  the  result for the 3D XY-model  (corresponding to the use of (\ref{eq:Hamiltonian3})) one finds  $T_N\approx2.2J/k_B$ \cite{Janke97}. 
  This approximation is restricted to a region  not too close to the Lifshitz point $\theta=0$.  
  Above $T_R$ the Hubert wall exhibits short range elasticity. However, since $T_R$ is of the order $T_N$, the non-local elasticity dominates over a large temperature region.

\smallskip

 I have also studied the possibility of   {\it disorder driven} roughening  of Hubert walls which would render its elasticity local \cite{Nattermann85}. 
 Using arguments similar to those used in \cite{Grinstein+82} I found that the effective step energy  surrounding an  terrace of linear size $L$ vanishes  on scales $L>a\exp[1/(|\theta|^7c_{\textrm{imp}})]$. Here 
 \eqa\label{impuritystrength}
 c_{\textrm{imp}}=n_{\textrm{imp}}v_0^2/a^3
 \eqe
 is a dimensional expression for the strength of collective pinning.   $n_{\textrm{imp}}$ and $v_0$ denote the impurity concentration and volume, respectively.   Hence
 terraces {\it are}  generated spontaneously by the disorder. Because of the   exponentially large  length scale  this effect will hardly be seen and hence Hubert walls remain flat and elasticity non-local as long as $T<T_R$.

\section{ Elasticity of vortex walls} 
Next I consider a DW whose normal is tilted away by an angle $\alpha$ from the $x$-axis. 
Such a wall includes  vortices. It  consists of steps considered in the previous section which are separated by terraces of average width 
$
\ell=2\pi a/(\theta\tan\alpha)
$ (see Fig.2, right panel). 
The  surface energy $\sigma(\alpha)$  of the DW can then be written as  
\begin{equation}
\sigma(\alpha)\approx\left[\sigma_H+\varepsilon_{\textrm{int}}(\ell)\right]|\cos\alpha|+ \sigma_v|\sin\alpha|. 
\end{equation}
$\sigma_H=J\theta^3/(3a^2)$  and 
$\sigma_v=\varepsilon_s\theta/(2\pi a)$ are the  surface tension of the Hubert ($\alpha=0$) and the pure vortex wall ($\alpha=\pi/2$), respectively.
 $\varepsilon_{\textrm{int}}(\ell)$ describes the  step interaction. 
 At $T=0$, 
 $\varepsilon_{\textrm{int}}(\ell)\approx{\sigma_v}\exp\left(-q\ell\right)$, whereas at $T>0$ $\varepsilon_{\textrm{int}}(\ell)\sim T/\ell^2$ due to collisions of meandering steps \cite{Noziere01}.
To obtain the elastic constants of the DW one has to consider  an infinitesimal homogeneous  distortion 
 $\partial_{\eta}u\equiv\epsilon$ away from the plane perpendicular to the DW normal $\hat n$ \cite{Chaikin+95}.
This  changes the surface energy density by
\begin{equation}\label{surfacetension2}
\frac{\sigma(\alpha+\epsilon)}{\cos\epsilon}-\sigma(\alpha)\approx\sigma'(\alpha)\epsilon+
 \frac{1}{2}\left[{\sigma(\alpha)+\sigma''(\alpha)}\right]\epsilon^2.
\end{equation}
The linear term in $\epsilon$
vanishes  at finite temperatures since the roughening transition temperature  for the vicinal surface considered here vanishes.  This follows from the fact that  Hubert walls are structureless in the plane orthogonal to $\hat{x}$ and hence steps can meander freely, leading to a rough surface \cite{Noziere01}.
 A distortion $\partial_{z}u$ leads to a similar expression without  derivative terms since the $\sigma(\alpha)$  depends only on $\alpha$. 
 The total elastic energy for the vortex wall can therefore be written as
 \begin{equation}\label{elasticenergy2}
{\hat{\cal G}}_v^{-1}({\mathbf k})=\gamma(\alpha) k_\eta^2+\sigma(\alpha)k_z^2,
\end{equation} 
i.e. vortex walls exhibit conventional elasticity. 
Note that  
\eqa
\gamma(\alpha)=\sigma(\alpha)+\sigma''(\alpha)=\varepsilon''_\textrm{int}\cos\alpha-2\varepsilon'_\textrm{int}\sin\alpha\eqe
only depends on the vortex interaction which is small for small $\alpha$. Here
$\varepsilon'_\textrm{int}=d\varepsilon'_\textrm{int}[\ell(\alpha)]/d\alpha$ etc.

  A corresponding calculation for $\alpha\approx\pi/2$ is more difficult. The height $h$ of steps in  pure vortex walls can take any value, but steps cannot meander freely since vortex walls have a structure periodic  in the $\hat x$-direction. However,  the step energy  is small in this case since  vortices can almost freely slide against each other and   hence $T_R\ll J/k_B$.

\section{ DW pinning by impurities}
DWs can be pinned by impurities \cite{Haasen77}.
 The statistical pinning theory of DWs with local elasticity has been developed some time ago \cite{Nattermann+92}.    
 In the present context it will be applied to Hubert and vortex walls.  
 I assume the presence of  non-magnetic impurities  which dilute the system and hence contribute a term 
\begin{equation}
{\cal H}_{\textrm{imp}}=-J\int_\br \delta\tau(\br){\cal E}[\bm\eta,u(\bm\eta)-\xi]\end{equation}
 to its energy. 
 Here $J{\cal E}$ is the energy density of the domain wall and $\delta\tau=v_0\sum_i\delta(\br-\br_i)$ where ${\mathbf r}_i$ denotes the impurity  position. 
 We will assume that  $v_0^{1/3}q\ll1$  such that the pinning energy correlation length parallel to $\xi$ is of the order of the DW width $\sim q^{-1}$. 
The local pinning force density  $f({\mathbf \eta},u)$ follows then from  
 \eqa 
 f=-\frac{\delta{\cal H}_{\textrm{imp}}}{\delta u(\bm\eta)}=\sum\nolimits_{i}J^2v_0\delta({\bm\eta}-{\bm \eta}_i){\mathcal E}'(\bm\eta,u-\xi_i),  \eqe 
where  ${\mathcal E}'(\bm\eta,u-\xi_i)=d {\mathcal E}(\bm\eta,u-\xi_i)/du$. Averaging over the random  $\br_i$  one obtains  $ \langle{f}\rangle=0$ and   
    \begin{equation} 
    \langle f({{\bm \eta}},u)f({{\bm \eta}}',u')\rangle=\delta({\bm \eta}-{\bm\eta}')\Delta_0(\bm\eta, u-u').
    \end{equation} 
Here $\langle...\rangle$ denotes the disorder average and 
\begin{equation}
\Delta_0(\bm\eta, u)=J^2c_{\textrm{imp}}a^3\int d\xi {\mathcal E}'(\bm\eta,u-\xi){\mathcal E}'(\bm\eta,-\xi).
\end{equation}  
  Second order perturbation theory gives for the pinning threshold of a driven DW \cite{Nattermann+92}
\begin{equation}\label{pinning-threshold}
f_{\textrm{pin}}(\bm\eta)=-\Delta_0'(\bm\eta,u\to\pm0) {\cal G}(0). 
\end{equation}
Here  $\pm$ \textrm{sign} denotes the \textrm{sign} of the driving force.
Generically  $\Delta_0'(\bm\eta,0)\sim-\int d\xi\Lk{\lK{\mathcal E}'(\bm\eta,u-\xi)\rK^2}\Rk'=0$ for  analytic $\Delta(u)$, since ${\mathcal E}'(x\to\pm\infty)=0$.  Then  $f_{\textrm{pin}}=0$, i.e.  there is {\it no} coercivity from  perturbation theory. %

A finite parameter renormalization group calculation gives a diverging coupling constant 
$
g_l=\Delta_l''(0), \,l=\ln(L/a),
$ 
when approaching  the (Larkin)  length scale ${\cal L}$. The latter follows from the balance between
  the elastic energy $E_{\textrm{el}}$ and the $E_{\textrm{pin}}\sim L^{D/2}$. 
If $E_{\textrm{el}}<E_{\textrm{pin}}$, the DW can adapt to the disorder and hence acommmodate to a potential valley where it gets  pinned. 
In the opposite case the DW is too stiff to stay in one valley. 
By crossing the rugged energy landscape,  potential forces on the  DW show either sign such that the resulting  pinning force  $\sim L$ and hence is surpassed by the driving force, which is typically $\sim L^2$.  
For systems with short range elasticity $E_{\textrm{el}}\sim L^{D-2}$. 
$D(=2)$ denotes the dimension of the domain wall. Hence  $E_{\textrm{el}}<E_{\textrm{pin}}$ for $D<D_c=4$ and $L>{\cal L}$, where $E_{\textrm{el}}({\cal L})=E_{\textrm{pin}}(\cal L)$. Since the random potential acting on the domain wall  is correlated in the $\xi$ direction over its width $q^{-1}$, the pinning force density can be estimated as 
\eqa\label{pinningforce}
f_{\textrm{pin}}\sim J q/{\cal L}^2.
\eqe

\section{Pinning of Hubert walls}
 As we have seen, Hubert walls exhibit non-local elasticity. Since the typical correlation length of the disorder seen by a domain wall is equal to its  width $\sim q^{-1}$, it makes sense to look at the elastic energy of a distortion $u\approx q^{-1}$ on the scale $L$.  
 This gives  with (\ref{elasticenergy2}) $
 E_{\textrm{el,H}}\sim JLq\sim J\theta L/a.
$ 
The variance of the pinning energy $E_\textrm{pin}$ is 
 \eqa\label{variance-pinning}
 \lgl {\cal H}_{\textrm{imp}}^2\rgl=(Jv_0)^2\sum\nolimits_i {\cal E}^2(\bm\eta_i,\xi_i)\approx (Jv_0)^2n_{\textrm{imp}}\int_{\xi,\bm\eta}{\cal E}^2.
 \eqe
The energy density of  planar  Hubert walls \cite{Hubert74} is obtained from (\ref{mmHamiltonian-phi1}) and (\ref{phi-0}) as
\eqa
{\cal E}_H(\bm\eta,\xi)\approx \frac{a}{2}\lK\phi_0''\lk q\xi\rk\rK^2= {\theta^4}/\lK{2a^3\cosh^4(q\xi)}\rK.
\eqe
We note that ${\cal E}_H$  depends  only  via $u$ on $\bm\eta$ in this cases. 
From the last two relations  
we get  for the Hubert wall 
\eqa\label{pinning-energy}
E_{\textrm{pin,H}}=c_HJc^{1/2}_{\textrm{imp}}\theta^{7/2}L/a
 \eqe
where $c_H\approx 0.49$. %
This gives for  the  ratio of the pinning  to the elastic energy
$E_\textrm{pin,H}/E_{\textrm{el,H}}\sim (\theta^5 c_{\textrm{imp}})^{1/2}\ll 1$. 
From this argument on does not expect any pinning of Hubert walls by impurities. 
However the argument presented did not take into account that independent pinning energy gains can be made on different length scales \cite{Grinstein+82}, which leads to an additional  logarithmic factor $\ln(Lq)$ in the pinning energy. As a result, $E_\textrm{pin,H}/E_{\textrm{el,H}}$ is of  order one on the exponentially large length scale 
\eqa\label{Larkin-length}
{\cal L}_H\sim q^{-1}\exp[1/(\theta^5 c_{\textrm{imp}})^{1/2}].
\eqe

\smallskip

\section{Functional renormalization group} We will now calculate the pinning force on a Hubert wall using the functional renormalization group calculation at the critical dimension $D_c=2$.
Following the calculation scheme used in \cite{Nattermann+92} one can show that the effective force correlator $\Delta_l(u)$ on scale $L=ae^l$ obeys the RG flow equation ($c=1/(64\pi J^2q^6)$)
\begin{align}\label{flow}
\!\frac{d\Delta_{l}(u)}{dl}=c
 \frac{d^2}{du^2}\Delta_l(u)\left[2\Delta_l(0)-\Delta_l(u)\right].
 \end{align}
Differentiation twice with respect to $u$     
one finds from (\ref{flow}), assuming $\Delta'(0)=0$, that $g$ obeys the equation 
\eqa
dg_l/dl=-6cg_l^2.
\eqe 
Integration with  the generic initial value $g_0 <0$  shows, that  $g$ develops a pole on      scale ${\mathcal L}_H=a\exp[1/(6cg_0)]$. 
On larger scale $\Delta_l(u)$ exhibits a {\it cusp} singularity. 
In this region (\ref{flow}) can  be solved with the ansatz 
\eqa
\Delta_l(u)=c^{-1}A^{2}l^{-1+2\mu}\Delta^*(uA^{-1}l^{-\mu}).
\eqe
This gives for $\Delta^*(u)$ the relation
\begin{equation} (1-2\mu-{\Delta^*}'')\Delta^*-\mu u{\Delta^*}'-{{\Delta^*}'}^2+{\Delta^*}''=0.\end{equation} 
I have chosen $A$ such that $\Delta^*(0)=1$. 
The value $\mu=1/3$ can be found from the fact that $\partial_l\int du\Delta_l(u)=0$. 
For small $u$ one then gets 
\eqa
\Delta^*(u)=1-|u|/\sqrt{3} + 2u^2/9.
\eqe
Thus 
$
{\Delta^*}'(\pm0)\sim-\textrm{sign}\,{u}
$
 and  $\infty>g^*>0$.
One can now apply  (\ref{pinning-threshold}) using the {\it renormalized} function $\Delta_l(u)$ on scales larger ${\mathcal L}_H$.  
${\mathcal L}_H$ plays  the role of the short length scale cut-off of the renormalized theory.  
With $g_0\approx-2.4c_{\textrm{imp}}J^2\theta^{11}a^{-6}$,  one obtains finally for the coercive force
 \begin{equation}
 \label{pinning_Hubert}
 |f^{(H)}_{\textrm{pin}}|\approx 
\lk{4|\theta|/a}\rk^3\pi J \exp\lK-c_p/(|\theta|^5 c_{\textrm{imp}})\rK.
\end{equation}  
  and $c_p\approx{14.2}$. This result is in  agreement
  with our previous estimate (\ref{pinningforce}), (\ref{Larkin-length}).
Since both $\theta,\,c_{\textrm{imp}}\ll{1}$, 
pinning of Hubert walls by direct interaction with impurities is {\it completely negligible}, as long as one is below $T_R$. 
This is a  direct consequence of their non-local elasticity. 
It   implies that after a quench to  a metastable multi-domain state Hubert walls will quickly disappear from the sample. 
Indeed, in films of helical magnets with  film plane perpendicular to the helical axis, domains were found to extend over the whole film width \cite{Weschke+04}.

\section{ Pinning of vortex walls} 
We come now to the consideration of the pining of vortex walls whose normal is tilted away from the $x$ direction by an angle $\alpha$. Since their elasticity is  short range, (see eq.(\ref{elasticenergy2}),) 
the elastic energy of a distortion $u\sim q^{-1}$ is of the order 
\eqa
E_{\textrm{el}}\approx q^{-2}\lK\sigma(\alpha)\gamma(\alpha)\rK^{1/2}.
\eqe 

To find the variance of the pinning energy of the steps  (e.g. in a wall where $\alpha=\pi/2$) 
 we use the energy density expression (\ref{energydensity2}) of the variational study of vortices  \cite{Schenck+13} which gives  for the pinning  energy
\eqa
E_{\textrm{pin,s}}\approx c_s J c^{1/2}_{\textrm{imp}}L/a
\eqe
and $c_s\approx 0.42$. The pining energy is dominated by contributions from the centre of the vortex and hence does not dependent on $\theta$.

We estimate the pinning energy of a wall of { general orientation}  
by
\eqa
E_{\textrm{pin}}\approx \lK\cos^2\!\alpha E^2_{\textrm{pin,H}} +\sin^2\!\alpha E^2_{\textrm{pin,s}}\rK^{1/2}\qquad\qquad\\\qquad =J c^{1/2}_{\textrm{imp}}c_s\frac{L}{a} \sin\!\alpha 
\lK1+\lk{ c_H\cot\!\alpha}/{c_s}\rk^2\theta^7\rK^{1/2}.
\eqe
Thus, as soon  as $\tan\alpha>c_H\theta^{7/2}/c_s$ pinning of the domain wall is dominated by the steps. In this region 
the Larkin length is then given by
\eqa
{\cal L}(\alpha)\sim{a^3\lK\sigma(\alpha)\gamma(\alpha)\rK^{1/2}}\lK{Jc^{1/2}_{\textrm{imp}}c_s\theta^2\sin\alpha}\rK^{-1}
\eqe
and correspondingly the pinning force density  (\ref{pinningforce}). 
Note, that this result does not cross-over to ${\cal L}_H$ since we assumed here short range elasticity. The latter is present only on scales $L\gg \ell(\alpha)$.
This result is confirmed by a functional renormalisation group calculation in $D=4-\epsilon$ dimensions with $\epsilon=2$ which increases the estimate of the Larkin length by a factor $70$.

\section{Anisotropy} 
In systems where the $U(1)$ symmetry of the magnetic structure is broken   Hubert walls are also pinned by  weak anisotropy,
as I will show now. It is easy to assure oneself that moving a rigid Hubert wall requires the rotation of {\it all} spins in at least one half space.  
This rotation does not cost energy even in the presence of impurities, as long as    the $U(1)$ symmetry is preserved and the interaction with impurities does not depend on the spin direction, as I assume here.
Most of the experimental systems have however a weak anisotropy of the form
\begin{equation}\label{anisotropy}
{\cal H}_{{v}}=-J({v}/{a^3})\int_{{\mathbf r}}\cos (p\phi),\quad{v}>0.
\end{equation} 
For sufficiently large anisotropy helical  regions  of width $w=a\theta/(pv)^{1/2}$ are separated by commensurate regions of almost constant phase $\phi=2\pi n/p$ \cite{Chaikin+95}. The latter increase with increasing anisotropy until at  $w<w_c\sim q^{-1}$ the system becomes ferromagnetic. Accommodation of a DW inside a commensurate region saves energy
and hence leads to pinning.

This effect works also  in the case of very weak anisotropy.  The ground state can then be described by  a weak modulation of the wave vector
\begin{equation} 
q(x)
\approx \pm q\Lk1-\nu\sin [pq(x-x_0)]\Rk,\quad \nu\sim{v}/{\theta^4}\ll{1}.
\end{equation}
The absolute value of the phase gradient $q(x)$ is smallest for $pq(x_n-x_0)\approx 2\pi n, n$ integer, corresponding to the almost commensurate regions for large anisotropy.  Assuming for the moment $\nu\approx 1$ $\phi_x$ reaches zero at $x=x_n$ for both helicities  and hence a domain wall would not cost extra energy. It is clear that this effect also works for $\nu<1$. Thus Hubert walls are pinned by the anisotropy.

\section{Non-centrosymmetric systems} 

In the generic case the Hamiltonian of non-centrosymmetric systems has the form \cite{Li+12}
 \begin{equation}\label{GL-NCS}
{\cal H}=\frac{J}{2}\int_{{\mathbf r}}\left[({\mathbf \nabla}{\mathbf m})^2+ |{\mathbf q}|{\mathbf m}({\mathbf \nabla}\times{\mathbf m})\right].
\end{equation} 
The direction of ${\mathbf q}=\theta\hat{q}/a$ is fixed by an additional   weak (cubic) anisotropy of the order $\theta^4$.  $\theta$ is here proportional to the  spin-orbit coupling. 
The anisotropy term  can therefore  be neglected otherwise.
Hubert walls are characterized  by a  normal $\hat{n}$ obeying     $\hat{n}\cdot{\mathbf q}_+=\hat{n}\cdot{\mathbf q}_-$. ${\mathbf q}_+$, ${\mathbf q}_-$ are the wave vectors of the adjacent domains. 
To be specific I consider ${\mathbf q}_\pm=(q_x(x),q_y,0)$ with $q_x(x)$ changing smoothly from $q_x$ to $-q_x$ over a region of size $q^{-1}$ when crossing the wall \cite{Li+12} (compare Fig. 1, right panel). 
Expressing  ${\mathbf m}=\hat{y}\cos\phi+\hat q(x)\times\hat{y}\sin\phi$, and ignoring terms which are non-zero only inside the Hubert wall (and hence do not contribute to non-local elasticity), one can rewrite (\ref{GL-NCS}) as \begin{equation}{\cal H}_0\approx({J}/{2})\int_{{\mathbf r}}\left[{\mathbf \nabla}\phi-{\mathbf q}(x)\right],^2\end{equation} which has    the same form as (\ref{eq:Hamiltonian3}).
Thus,  Hubert walls in non-centrosymmetric systems show long range elasticity as well. All other conclusions made for Hubert wall and vortex walls transfer straightforwardly.

\section{ Liquid crystals}Chiral nematic and smectic phases of liquid crystals exhibit  helical phase \cite{Kleman70,Kleman+03}. Their description is similar to that used in the  non-centrosymmetric case, (\ref{GL-NCS}), provided ${\mathbf m}$ is replaced by the {\it director} $\bf{n}$,  and $J$ by the Frank constant.  
 Further, in contrast to helimagnets, the direction of ${\mathbf q}$ is not fixed in space by anisotropy since chirality is introduced through the chirality of the molecules. DWs of the type described above will occur however as grain boundaries between  phases with different ${\mathbf q}$ direction. A detailed discussion is   beyond the scope of the present paper.

\acknowledgments
  I thank R.D. Kamien,  A. Rosch, B. Roostaei and C. Sch\"ussler-Langeheine  for  useful discussions,  
  V.L Pokrovsky for a careful reading of the manuscript, and  the Sino-French Center, Sun Yat-sen University Guangzhou for hospitality. This
 work has been supported by SFB 608 of the DFG.
\bibliographystyle{eplbib}
\bibliography{Helimagnet_pinning}

\begin{thebibliography}{10}
\expandafter\ifx\csname url\endcsname\relax\def\url#1{\texttt{#1}}\fi

\bibitem{Li+12}
\Name{Li F., Nattermann T. \and Pokrovsky V.~L.} \REVIEW{Phys. Rev.
  Lett.}{108}{2012}{107203}.

\bibitem{Haasen77}
\Name{Haasen P.} \REVIEW{Contemp. Phys.}{18}{1977}{373}.

\bibitem{Frohlich54}
\Name{Fr{\"o}hlich H.} \REVIEW{Proc. R. Soc. London, Ser.A}{223}{1954}{296}.

\bibitem{Hubert74}
\Name{Hubert A. \and Sch\"afer R.} \Book{Magnetic domains: the analysis of
  magnetic microstructures} (Springer) 2009.

\bibitem{Kleemann07}
\Name{Kleemann W.} \REVIEW{Annu. Rev. Mater. Sci.}{37}{2007}{415}.

\bibitem{Parkin+08}
\Name{Parkin S. S.~P., Hayashi M. \and Thomas L.}
  \REVIEW{Science}{320}{2008}{190}.

\bibitem{Brazovskii+04}
\Name{Brazovskii S. \and Nattermann T.} \REVIEW{Adv. Phys.}{53}{2004}{177}.

\bibitem{Jensen+91}
\Name{Jensen J. \and Mackintosh A.} \Book{Rare Earth Magnetism Structures and
  Excitations} (Oxford UP, New York) 1991.

\bibitem{Lang+04}
\Name{Lang J.~C., Lee D.~R., Haskel D. \and Srajer G.} \REVIEW{J. Appl.
  Phys.}{95}{2004}{6537}.

\bibitem{Pfleiderer+04}
\Name{Pfleiderer C. \etal} \REVIEW{Nature}{427}{2004}{227}.

\bibitem{Uchida+06}
\Name{Uchida M., Onose Y., Matsui Y. \and Tokura Y.}
  \REVIEW{Science}{311}{2006}{359}.

\bibitem{Uchida+08}
\Name{Uchida M. \etal} \REVIEW{Phys. Rev. B}{77}{2008}{184402}.

\bibitem{Kim+Seik+08}
\Name{Kimura T. \etal} \REVIEW{Nature Mat.}{7}{2008}{291}.

\bibitem{Mostovoy06}
\Name{Mostovoy M.} \REVIEW{Phys. Rev. Lett.}{96}{2006}{067601}.

\bibitem{Cheong+07}
\Name{Cheong S.-W. \and Mostovoy M.} \REVIEW{Nature Mat.}{6}{2007}{13}.

\bibitem{Chapon+04}
\Name{Chapon L.~C. \etal} \REVIEW{Phys. Rev. Lett.}{93}{2004}{177402}.

\bibitem{Kimura+08}
\Name{Kimura T. \and Tokura Y.} \REVIEW{J. Phys. Cond. Mat.}{20}{2008}{434204}.

\bibitem{Tal+08}
\Name{Talbayev D. \etal} \REVIEW{Phys. Rev. Lett.}{101}{2008}{247601}.

\bibitem{Yam+06}
\Name{Yamasaki Y. \etal} \REVIEW{Phys. Rev. Lett.}{96}{2006}{207204}.

\bibitem{Chap+11}
\Name{Chapon L.~C. \etal} \REVIEW{Phys. Rev. B}{83}{2011}{024409}.

\bibitem{Arima11}
\Name{Arima T.} \REVIEW{J. Phys. Soc. Jap.}{80}{2011}{052001}.

\bibitem{Weschke+04}
\Name{Weschke E. \etal} \REVIEW{Phys. Rev. Lett.}{93}{2004}{157204}.

\bibitem{Andrianov+00}
\Name{Andrianov A.~V., Kosarev D.~I. \and Beskrovnyi A.~I.} \REVIEW{Phys. Rev.
  B}{62}{2000}{13844}.

\bibitem{deGennes62}
\Name{Gennes P.~D.} \REVIEW{J. Phys. Radium}{23}{1962}{}.

\bibitem{Joanny+84}
\Name{Joanny J.~F. \and de~Gennes P.~G.} \REVIEW{J. Chem.
  Phys.}{81}{1984}{552}.

\bibitem{Chaikin+95}
\Name{Chaikin P. \and Lubensky T.} \Book{Principles of Condensed Matter
  Physics} (Cambridge UP) 1995.

\bibitem{Noziere01}
\Name{Noziere P.} \REVIEW{Eur. Phys. J. B}{24}{2001}{383}.

\bibitem{Schenck+13}
\Name{Schenck H., Pokrovsky V. \and Nattermann T.} \REVIEW{arXiv:1308.0823
  [cond-mat.stat-mech]}{}{2013}{}.

\bibitem{Hasenbusch+97}
\Name{Hasenbusch M. \and Pinn K.} \REVIEW{J. Phys. A: Math.
  Gen.}{30}{1997}{63}.

\bibitem{Janke97}
\Name{Janke W.} \REVIEW{Phys. Lett. A}{148}{1997}{306}.

\bibitem{Nattermann85}
\Name{Nattermann T.} \REVIEW{phys. status solidi (b)}{132}{1985}{125}.

\bibitem{Grinstein+82}
\Name{Grinstein G. \and Ma S.} \REVIEW{Phys. Rev. Lett.}{49}{1982}{685}.

\bibitem{Nattermann+92}
\Name{Nattermann T. \etal} \REVIEW{J. de Physique II}{2}{1992}{1483}.

\bibitem{Kleman70}
\Name{Kl\'eman M.} \REVIEW{Phil. Mag.}{22}{1970}{739}.

\bibitem{Kleman+03}
\Name{Kl\'eman M. \and Lavrentovicht O.~D.} \Book{Soft Matter Physics: An
  Introduction} (Springer, New York) 2003.

\end{thebibliography}
\end{document}